\documentclass[preprin,showpacs,preprintnumbers,eqsecnum,amsmath,amssymb]{revtex4}
\usepackage{graphics,epsfig}
\usepackage{graphicx}
\usepackage{dcolumn}
\usepackage{bm}

\begin{document}
\baselineskip=0.8 cm
\title{{\bf Strong field gravitational lensing in the deformed
Ho\v{r}ava-Lifshitz black hole }}

\author{Songbai Chen}
\email{csb3752@163.com} \affiliation{Institute of Physics and
Department of Physics,
Hunan Normal University,  Changsha, Hunan 410081, P. R. China \\
Key Laboratory of Low Dimensional Quantum Structures and Quantum
Control (Hunan Normal University), Ministry of Education, P. R.
China.}

\author{Jiliang Jing }
\email{jljing@hunnu.edu.cn}
 \affiliation{Institute of Physics and
Department of Physics,
Hunan Normal University,  Changsha, Hunan 410081, P. R. China \\
Key Laboratory of Low Dimensional Quantum Structures and Quantum
Control (Hunan Normal University), Ministry of Education, P. R.
China.}

\vspace*{0.2cm}
\begin{abstract}
\baselineskip=0.6 cm
\begin{center}
{\bf Abstract}
\end{center}

Adopting the strong field limit approach, we studied the properties
of strong field gravitational lensing  in the deformed
Ho\v{r}ava-Lifshitz black hole and obtained the angular position and
magnification of the relativistic images. Supposing that the
gravitational field of the supermassive central object of the galaxy
described by this metric, we estimated the numerical values of the
coefficients and observables for gravitational lensing in the strong
field limit. Comparing with the Reissner-Norstr\"{om} black hole, we
find that with the increase of parameter $\alpha$, the angular
position $\theta_{\infty}$ decreases more slowly and $r_m$ more
quickly, but angular separation $s$ increases more rapidly. This may
offer a way to distinguish a deformed Ho\v{r}ava-Lifshitz black hole
from a Reissner-Norstr\"{om} black hole by the astronomical
instruments in the future.
\end{abstract}

 \maketitle
\newpage
\vspace*{0.2cm}
\section{Introduction}

The general theory of relativity tells us that photons would be
deviated from their straight path when they pass close to a compact
and massive body. The phenomena resulting from the deflection of
light rays in a gravitational field are called as gravitational
lensing and the object causing a detectable deflection is usually
named as a gravitational lens. Like a natural and large telescope,
gravitational lensing can provide us the information about the
distant stars which are otherwise too dim to be observed. Moreover,
it can help us to detect the exotic objects (such as cosmic strings)
in the universe as well as to verify alternative theories of
gravity. Most of the theories of gravitational lensing have been
developed in the weak field approximation
\cite{Schneider}-\cite{RDB} in which one only keeps the first non
null term in the expansion of the deflection angle. In general, it
is enough for us to investigate the properties of gravitational
lensing by ordinary stars and galaxies. However, when the lens is a
compact object with a photon sphere (such as black hole), a strong
field treatment of gravitational lensing
\cite{Darwin,Vir,Vir1,Vir2,Vir3,Fritt} is need instead because that
in this situation photons passing close to the photon sphere have
large deflection angles and the weak field approximation is no valid
any more. Virbhadra and Ellis \cite{Vir1} find that near the line
connecting the source and the lens, an observer would detect two
infinite sets of faint relativistic images on each side of the black
hole which are produced by photons that make complete loops around
the black hole before reaching the observer. These relativistic
images could provide a profound verification of alternative theories
of gravity in their strong field regime. Thus, the study of the
strong field gravitational lensing by black holes in the different
spacetimes becomes appealing recent years.

On the basis of the Virbhadra-Ellis lens equation \cite{Vir2,Vir3},
Bozza \cite{Bozza1} devised an analytical method for calculating the
deflection angles for the light rays propagating close to the
Schwarzschild black hole and showed that there exists a logarithmic
divergence of the deflection angles at photon sphere. Later Bozza's
technique was extended to other static spacetimes. For example,
Eiroa \textit{et al} \cite{Eirc1}-\cite{Eirc3} have studied the
gravitational lensing due to the Reissner-Nordstr\"{o}m black hole,
braneworld black hole and Einstein-Born-Infeld black hole. Bozza
\cite{Bozza2} extended the analytical method of lensing for a
general class of static and spherically symmetric spacetimes and
showed that the logarithmic divergence of the deflection angle at
photon sphere is a common feature for such spacetimes. Moreover, he
\cite{Bozza3}-\cite{Bozza5} has also studied the gravitational
lensing by a spinning black hole. Bhadra \textit{et al}
\cite{Bhad1}\cite{Sarkar} have considered the gravitational lensing
by the Gibbons-Maeda-Garfinkle-Horowitz-Strominger charged black
hole and the black hole in the Brans-Dicke theory. Konoplya
\cite{Konoplya1} has studied the corrections to the deflection angle
and time delay in the background of a black hole immersed in a
uniform magnetic field. Majumdar \cite{Muk} has investigated the
gravitational lensing in the dilaton-de Sitter black hole
spacetimes. Perlick \cite{Per} has obtained an exact gravitational
lens equation in a spherically symmetric and static spacetime and
used it to study lensing by a Barriola-Vilenkin monopole black hole.
Gyulchev \cite{GNG} has studied the features of light propagation
close to the equatorial plane of the rotating dilaton-axion black
hole spacetime and obtained that there exists a significant
dilaton-axion effect present on the gravitational lensing
observables in the strong field limit. These results are very useful
for us to verify the validity of gravity theories through the
astronomical observation of the relativistic images in the universe.

Recently, Ho\v{r}ava \cite{ho1} proposes a renormalizable
four-dimensional theory of gravity, which admits the Lifshitz
scale-invariance in time and space rather than Lorentz invariant
theory of gravity in $3+1$ dimensions. Thereafter, the
Ho\v{r}ava-Lifshitz gravity theory has been intensively investigated
in \cite{ho2,ho3,VW,klu,Nik,Nas,Iza,Vol,CH,CHZ} and its cosmological
applications have been  studied in \cite{cal,TS,muk,Bra,pia,gao,KK}.
Some static spherically symmetric black hole solutions have been
found in Ho\v{r}ava-Lifshitz theory \cite{CY,KS,LMP,CCO,CLS,Gho} and
the associated thermodynamic properties with those black hole
solutions have been investigated in \cite{MK,Nis,CCO1,Myung}. The
quasinormal modes of the massless scalar perturbations
\cite{sb1,RAK1} and the gravity lens in the weak field limit
\cite{RAK1} have been studied in the deformed Ho\v{r}ava-Lifshitz
black hole spacetime. Since the weak field limit takes the first
order deviation from Minkowski spacetime, it is necessary to study
the gravity lens in the strong field limit in the black hole
spacetime because that it starts from complete capture of the photon
and takes the leading order in the divergence of the deflection
angle. The main purpose of this paper is to study the gravity lens
in the strong field limit in the deformed Ho\v{r}ava-Lifshitz black
hole spacetime \cite{KS}.

The plan of our paper is organized as follows. In Sec.II we adopt to
Bozza's method and obtain the deflection angles for light rays
propagating in the deformed Ho\v{r}ava-Lifshitz black hole. In
Sec.III we suppose that the gravitational field of the supermassive
black hole at the centre of our galaxy can be described by this
metric and then obtain the numerical results for the observational
gravitational lensing parameters defined in Sec.II. Then, we make a
comparison between the properties of gravitational lensing in the
deformed Ho\v{r}ava-Lifshitz and Reissner-Norstr\"{om} metrics. At
last, we present a summary.

\section{Deflection angle in the deformed
Ho\v{r}ava-Lifshitz black hole}

In the Horava-Lifshitz gravity, the deformed action in the limit
$\Lambda_W \to 0$ can be described by \cite{KS}
\begin{eqnarray}
S_{HL}&=&\int dtd^3x \Big({\cal L}_0 + \tilde{{\cal L}}_1\Big),\\
{\cal L}_0 &=& \sqrt{g}N\left\{\frac{2}{\kappa^2}(K_{ij}K^{ij}
\label{action1}-\lambda K^2)+\frac{\kappa^2\mu^2(\Lambda_W R
  -3\Lambda_W^2)}{8(1-3\lambda)}\right\}\,,\\ \tilde{{\cal L}}_1&=&
\sqrt{g}N\left\{\frac{\kappa^2\mu^2 (1-4\lambda)}{32(1-3\lambda)}R^2
-\frac{\kappa^2}{2w^4} \left(C_{ij} -\frac{\mu w^2}{2}R_{ij}\right)
\left(C^{ij} -\frac{\mu w^2}{2}R^{ij}\right) +\mu^4R
\right\}.\label{action2}
\end{eqnarray}
Here $w$, $\lambda$, $\mu$ and $\kappa$ are the parameters in the
Horava-Lifshitz theory. $K_{ij}$ is extrinsic curvature
\begin{eqnarray}
K_{ij} = \frac{1}{2N} \Bigg(\partial_t g_{ij} - \nabla_i N_j -
\nabla_j N_i\Bigg).
\end{eqnarray}
and $C_{ij}$ is the Cotton tensor
\begin{eqnarray}
C^{ij}=\epsilon^{ik\ell}\nabla_k\left(R^j{}_\ell
-\frac14R\delta_\ell^j\right)=\epsilon^{ik\ell}\nabla_k R^j{}_\ell
-\frac14\epsilon^{ikj}\partial_kR,\label{def.K.C}
\end{eqnarray}
respectively. For $\lambda=1$, there exist a static and
asymptotically flat black hole solution which has a form \cite{KS}
\begin{eqnarray}
ds^2 = -f(r)\,dt^2 + \frac{dr^2}{f(r)} + r^2 (d\theta^2
+\sin^2\theta d\phi^2)\,, \label{metr}
\end{eqnarray}
and
\begin{eqnarray}
\label{sol1}
f(r)=\frac{2(r^2-2Mr+\alpha)}{r^2+2\alpha+\sqrt{r^4+8\alpha Mr}},
\end{eqnarray}
where $\alpha=1/(2w)$ and $M$ is an integration constant related to
the mass. Obviously, it returns the Schwarzschild spacetime as the
parameter $\alpha=0$. When the mass $M=0$, it is corresponding to
the Minkowski spacetime. Although the metric of this black hole
behaviors as that of Reissner-Norstr\"{om} black hole and the event
horizons are given by
\begin{eqnarray}
r_\pm=M\pm \sqrt{M^2-\alpha},
\end{eqnarray}
there exist a distinct difference between them is that the
denominator of $f(r)$ in the deformed Ho\v{r}ava-Lifshitz black hole
metric is no longer equal to $r$, which will make a great deal
influence on gravitational lensing in the strong field limit.

As in \cite{Vir2,Vir3,Bozza2}, we just consider that both the
observer and the source lie in the equatorial plane in in the
deformed Ho\v{r}ava-Lifshitz black hole (\ref{metr}) and the whole
trajectory of the photon is limited on the same plane. With the
conditions that $\theta=\frac{\pi}{2}$ and $2M=1$, the metric
(\ref{metr}) can be rewritten as
\begin{eqnarray}
ds^2=-A(r)dt^2+B(r)dr^2+C(r)d\phi^2,\label{grm}
\end{eqnarray}
with
\begin{eqnarray}
A(r)=f(r), \;\;\;\;B(r)&=&1/f(r),\;\;\;\; C(r)=r^2.
\end{eqnarray}
The deflection angle for the photon coming from infinite can be
expressed as
\begin{eqnarray}
\alpha(r_0)=I(r_0)-\pi,
\end{eqnarray}
where $r_0$ is the closest approach distance and $I(r_0)$ is
\cite{Vir2,Vir3}
\begin{eqnarray}
I(r_0)=2\int^{\infty}_{r_0}\frac{\sqrt{B(r)}dr}{\sqrt{C(r)}
\sqrt{\frac{C(r)A(r_0)}{C(r_0)A(r)}-1}}.\label{int1}
\end{eqnarray}
It is easy to obtain that as parameter $r_0$ decrease the
deflection angle increase. At certain a point, the deflection
angle will become $2\pi$, it means that the light ray will make a
complete loop around the compact object before reaching the
observer. When $r_0$ is equal to radius of the photon sphere, the
deflection angle diverges and the photon is captured.

The photon sphere equation is given by \cite{Vir2,Vir3}
\begin{eqnarray}
\frac{C'(r)}{C(r)}=\frac{A'(r)}{A(r)},\label{root}
\end{eqnarray}
which admits at least one positive solution and then the largest
real root of Eq.(\ref{root}) is defined  as the radius of the photon
sphere. In the deformed Ho\v{r}ava-Lifshitz black hole metric, the
radius of the photon sphere can be given explicitly by
\begin{eqnarray}
r_{ps}=\frac{3+(\sqrt{256\alpha^2-27}-16\alpha^2)^{2/3}}{2(\sqrt{256\alpha^2-27}-16\alpha^2)^{1/3}}.
\end{eqnarray}
When $\alpha=0$, it can recovers that in the Schwarzschild black
hole spacetime in which $r_{ps}=1.5$. However, it is quite a
different from that in the Reissner-Norstr\"{om} black hole
spacetime, which implies that there exist some distinct effects of
the Ho\v{r}ava-Lifshitz parameter $\alpha$ on gravitational lensing
in the strong field limit. Following the method developed by Bozza
\cite{Bozza2}, we define a variable
\begin{eqnarray}
z=1-\frac{r_0}{r},
\end{eqnarray}
and obtain
\begin{eqnarray}
I(r_0)=\int^{1}_{0}R(z,r_0)f(z,r_0)dz,\label{in1}
\end{eqnarray}
where
\begin{eqnarray}
R(z,r_0)&=&\frac{2r^2\sqrt{A(r)B(r)C(r_0)}}{r_0C(r)}=2,
\end{eqnarray}
\begin{eqnarray}
f(z,r_0)&=&\frac{1}{\sqrt{A(r_0)-A(r)C(r_0)/C(r)}}.
\end{eqnarray}
The function $R(z, r_0)$ is regular for all values of $z$ and
$r_0$. However, $f(z, r_0)$ diverges as $z$ tends to zero. Thus,
we split the integral (\ref{in1}) into two parts
\begin{eqnarray}
I_D(r_0)&=&\int^{1}_{0}R(0,r_{ps})f_0(z,r_0)dz, \nonumber\\
I_R(r_0)&=&\int^{1}_{0}[R(z,r_0)f(z,r_0)-R(0,r_{ps})f_0(z,r_0)]dz
\label{intbr},
\end{eqnarray}
where $I_D(r_0)$ and $I_R(r_0)$ denote the divergent and regular
parts in the integral (\ref{in1}), respectively. To find the order
of divergence of the integrand, we expand the argument of the square
root in $f(z,r_0)$ to the second order in $z$ and obtain the
function $f_0(z,r_0)$:
\begin{eqnarray}
f_0(z,r_0)=\frac{1}{\sqrt{p(r_0)z+q(r_0)z^2}},
\end{eqnarray}
where
\begin{eqnarray}
p(r_0)&=& 2-\frac{3r_0}{\sqrt{r^4_0+4\alpha r_0}},  \nonumber\\
q(r_0)&=&\frac{3r_0(r^3_0+\alpha)}{(r^3_0+4\alpha)\sqrt{r^4_0+4\alpha
r_0}}-1.
\end{eqnarray}
When $r_0$ is equal to the radius of photon sphere $r_{ps}$, the
coefficient $p(r_0)$ vanishes and the leading term of the
divergence in $f_0(z,r_0)$ is $z^{-1}$, thus the integral
(\ref{in1}) diverges logarithmically. Close to the divergence,
Bozza \cite{Bozza2} found that the deflection angle can be
expanded in the form
\begin{eqnarray}
\alpha(\theta)=-\bar{a}\log{\bigg(\frac{\theta
D_{OL}}{u_{ps}}-1\bigg)}+\bar{b}+O(u-u_{ps}),
\end{eqnarray}
where
\begin{eqnarray}
&\bar{a}&=\frac{1}{\sqrt{q(r_{ps})}}, \nonumber\\
&\bar{b}&=
-\pi+b_R+\bar{a}\log{\frac{r^2_{ps}[C''(r_{ps})A(r_{ps})-C(r_{ps})A''(r_{ps})]}{u_{ps}
\sqrt{A^3(r_{ps})C(r_{ps})}}}, \nonumber\\
&b_R&=I_R(r_{ps}), \;\;\;\;\;u_{ps}=\frac{r_{ps}}{\sqrt{A(r_{ps})}}.
\end{eqnarray}
\begin{figure}[ht]
\begin{center}
\includegraphics[width=5.0cm]{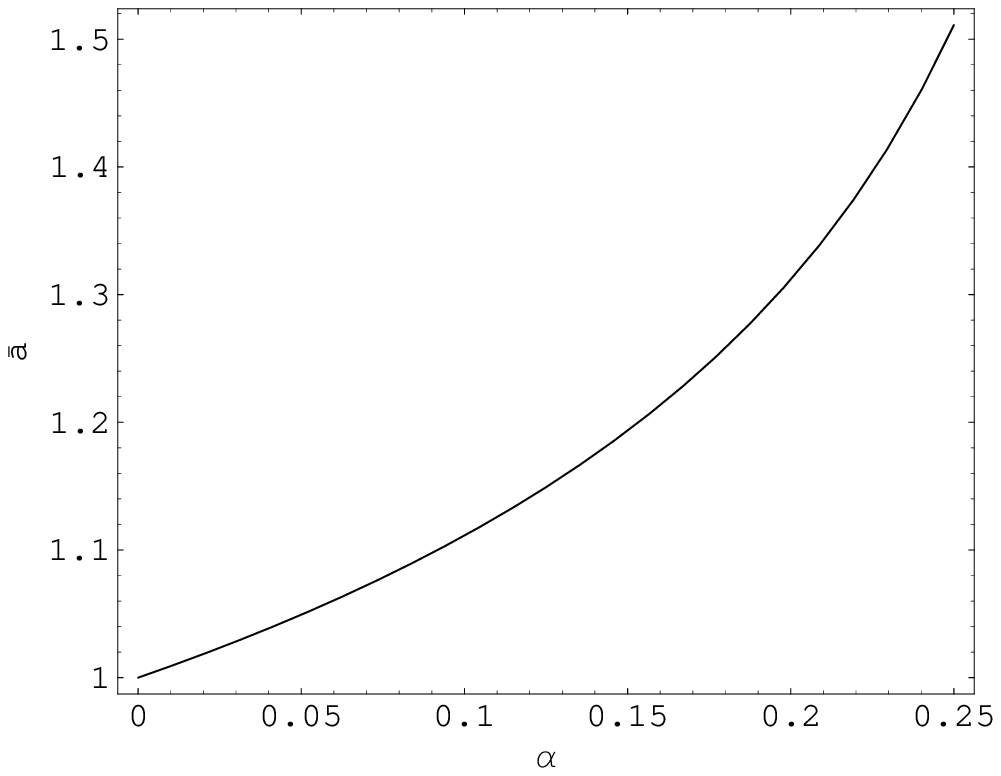}\;\;\;\;\;
 \includegraphics[width=5.0cm]{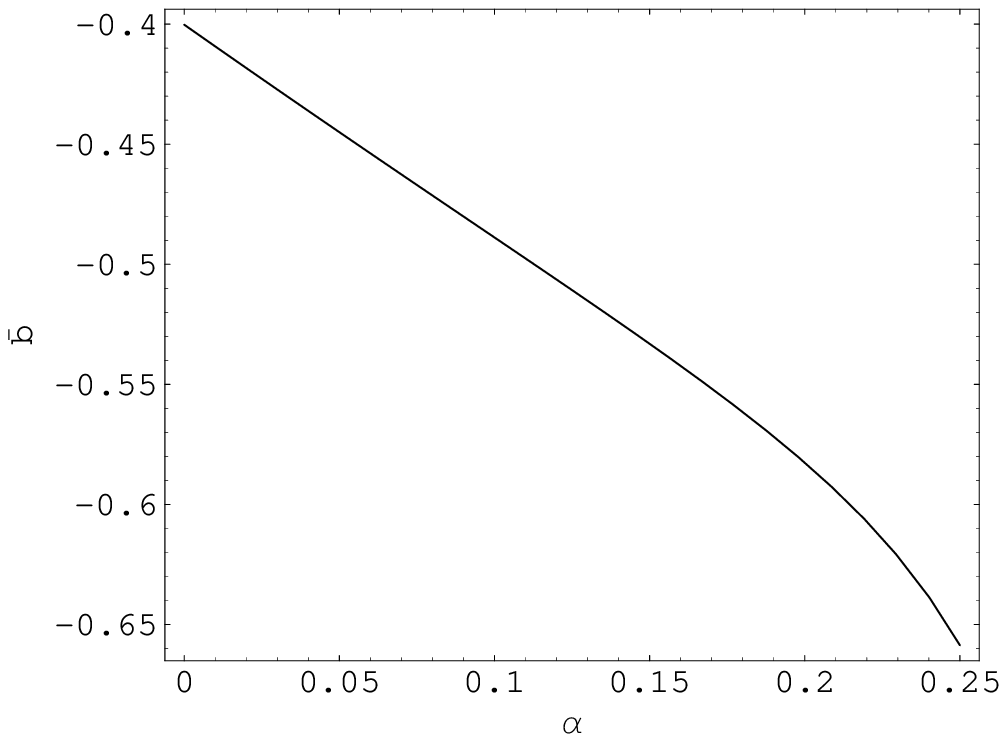}\;\;\;\;\includegraphics[width=5.0cm]{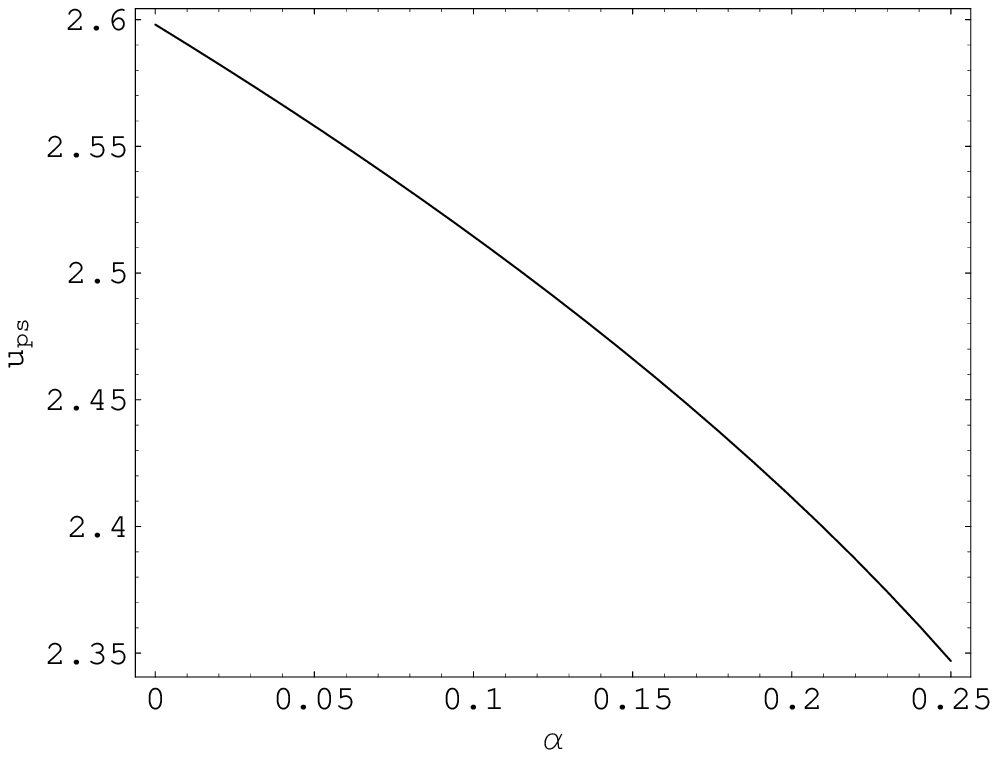}\\
 \includegraphics[width=5.0cm]{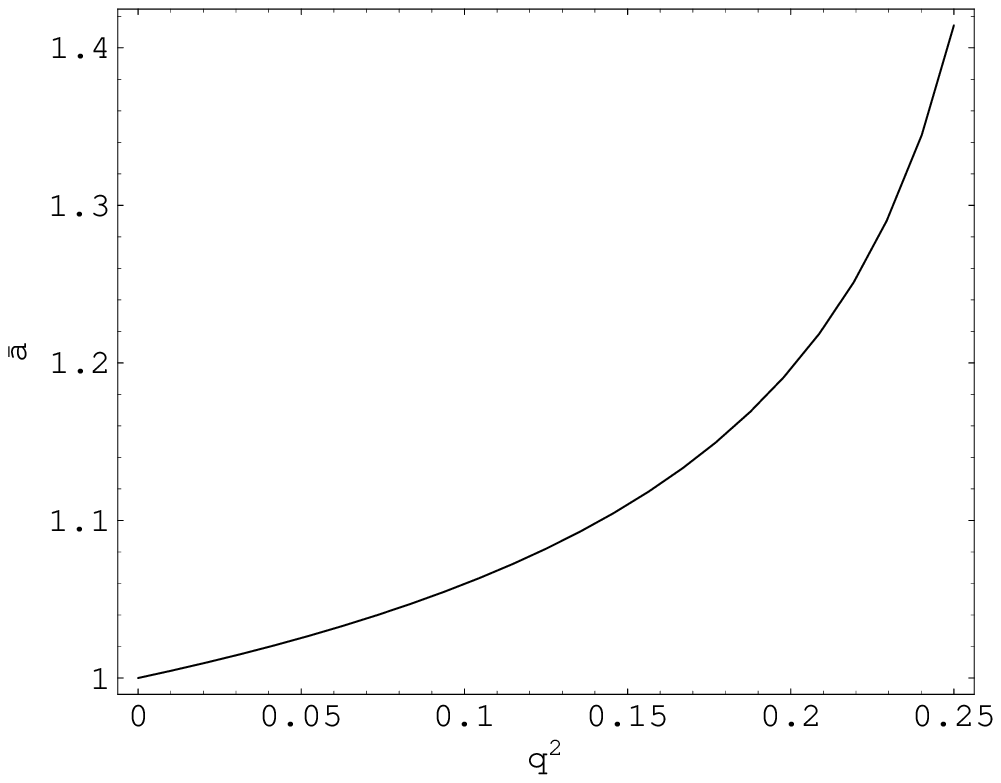}\;\;\;\;\;
 \includegraphics[width=5.0cm]{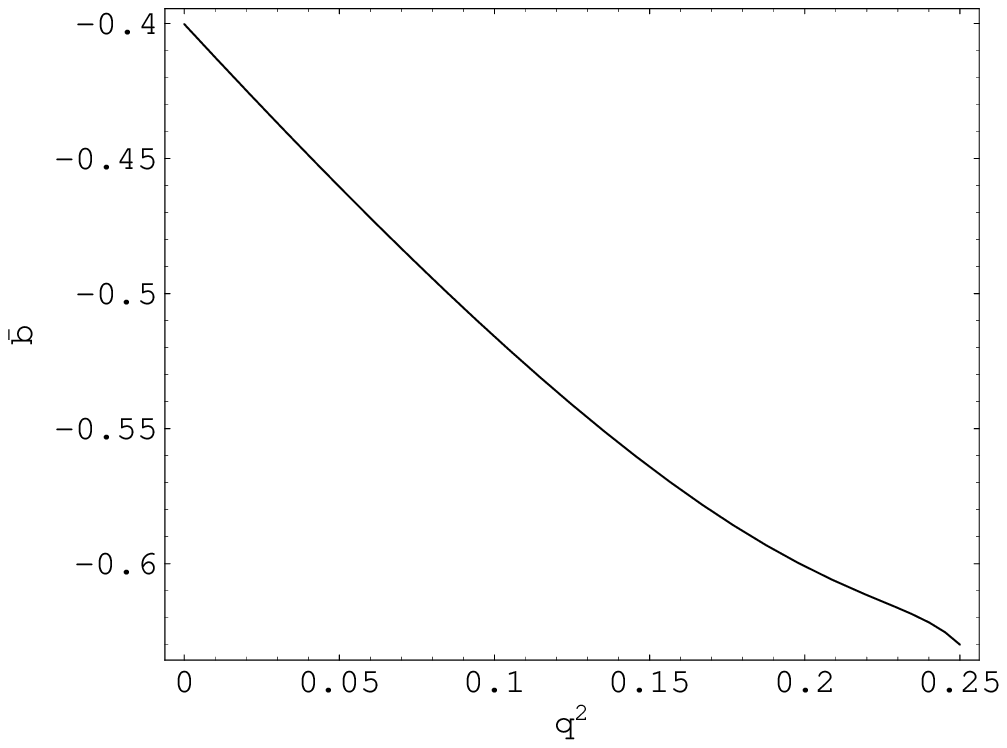}\;\;\;\;\includegraphics[width=5.0cm]{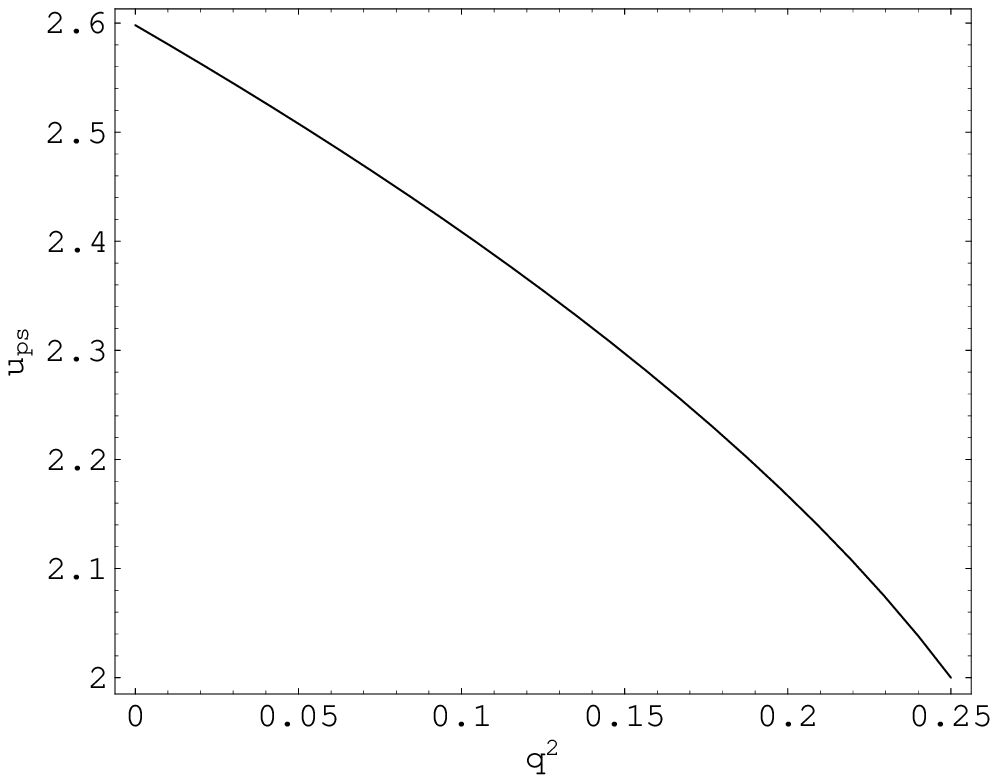}
\caption{ Variation of the coefficients of the strong field limit
$\bar{a}$, $\bar{b}$ and the minimum impact parameter $u_{ps}$ with
parameter $\alpha$ in the deformed Ho\v{r}ava-Lifshitz black hole
spacetime (in the upper row) and with $q^2$ in the
Reissner-Norstr\"{om} black hole (in the lower row).}
 \end{center}
 \label{f1}
 \end{figure}
$D_{OL}$ denotes the distance between observer and gravitational
lens, $\bar{a}$ and $\bar{b}$ are so-called the strong field limit
coefficients which depend on the metric functions evaluated at
$r_{ps}$. In general, the coefficient $b_R$ can not be calculated
analytically and need to be evaluated numerically. Here we expand
the integrand in (\ref{intbr}) in powers of $\alpha$ as in
\cite{Bozza2}, we can get
\begin{eqnarray}
b_R=b_{R,0}+b_{R,1}\alpha+ O(\alpha)^2,
\end{eqnarray}
and evaluate the single coefficients $b_{R,0}$ and $b_{R,1}$.
$b_{R,0}$ is the value of the coefficient for a Schwarzschild black
hole and $b_{R,1}$ is the correction from the Ho\v{r}ava-Lifshitz
parameter $\alpha$,
\begin{eqnarray}
b_{R,1}=\frac{16}{45}\bigg[-13+4\sqrt{3}+10\log{(3-\sqrt{3})}\bigg]=-1.3148,
\end{eqnarray}
which is larger than that of the Reissner-Norstr\"{om} black hole
where $b_{R,1}=-1.5939$ \cite{Bozza2}. Figures (1) tell us that with
the increase of $\alpha$ increases the coefficient $\bar{a}$
increase, but both of the minimum impact parameter $u_{ps}$ and
another coefficient $\bar{b}$ increases, which is similar to that in
in the Reissner-Norstr\"{om} black hole metric. However, as shown in
Fig. (1), in the deformed Ho\v{r}ava-Lifshitz black hole, $\bar{a}$
increases more quickly, both of $\bar{b}$ and $u_{ps}$ decrease more
slowly. This means that in principle we can distinguish a deformed
Ho\v{r}ava-Lifshitz black hole from a Reissner-Nordstrom black hole
by using strong field gravitational lensing.

Considering the source, lens and observer are highly aligned, the
lens equation in strong gravitational lensing can be written as
\cite{Bozza1}
\begin{eqnarray}
\beta=\theta-\frac{D_{LS}}{D_{OS}}\Delta\alpha_{n},
\end{eqnarray}
where $D_{LS}$ is the distance between the lens and the source,
$D_{OS}=D_{LS}+D_{OL}$, $\beta$ is the angular separation between
the source and the lens, $\theta$ is the angular separation
between the imagine and the lens, $\Delta\alpha_{n}=\alpha-2n\pi$
is the offset of deflection angle and $n$ is an integer. The
position of the $n$-th relativistic image can be approximated as
\begin{eqnarray}
\theta_n=\theta^0_n+\frac{u_{ps}e_n(\beta-\theta^0_n)D_{OS}}{\bar{a}D_{LS}D_{OL}},
\end{eqnarray}
where
\begin{eqnarray}
e_n=e^{\frac{\bar{b}-2n\pi}{\bar{a}}},
\end{eqnarray}
$\theta^0_n$ are the image positions corresponding to
$\alpha=2n\pi$.  The magnification of $n$-th relativistic image is
given by
\begin{eqnarray}
\mu_n=\frac{u^2_{ps}e_n(1+e_n)D_{OS}}{\bar{a}\beta
D_{LS}D^2_{OL}}.
\end{eqnarray}
If $\theta_{\infty}$ represents the asymptotic position of a set
of images in the limit $n\rightarrow \infty$, the minimum impact
parameter $u_{ps}$ can be simply obtained as
\begin{eqnarray}
u_{ps}=D_{OL}\theta_{\infty}
\end{eqnarray}
In the simplest situation, we consider only that the outermost
image $\theta_1$ is resolved as a single image and all the
remaining ones are packed together at $\theta_{\infty}$. Then the
angular separation between the first image and other ones can be
expressed as
\begin{eqnarray}
s=\theta_1-\theta_{\infty},
\end{eqnarray}
and the ratio of the flux from the first image and those from the
all other images is given by
\begin{eqnarray}
\mathcal{R}=\frac{\mu_1}{\sum^{\infty}_{n=2}\mu_{n}}.
\end{eqnarray}
For highly aligned source, lens and observer geometry, these
observable can be simplified as
\begin{eqnarray}
&s&=\theta_{\infty}e^{\frac{\bar{b}-2\pi}{\bar{a}}},\nonumber\\
&\mathcal{R}&= e^{\frac{2\pi}{\bar{a}}}.
\end{eqnarray}
The strong deflection limit coefficients $\bar{a}$, $\bar{b}$ and
the minimum impact parameter $u_{ps}$ can be obtain through
measuring $s$, $\mathcal{R}$ and $\theta_{\infty}$. Then,
comparing their values with those predicted by the theoretical
models, we can identify the nature of the black hole lens.

\section{Numerical estimation of observational gravitational lensing parameters}

In this section, supposing that the gravitational field of the
supermassive black hole at the galactic center of Milk Way can be
described by the deformed Ho\v{r}ava-Lifshitz black hole spacetime,
we estimate the numerical values for the coefficients and
observables of gravitational lensing in the strong field limit, and
then we study the effect of the metric parameter $\alpha$ on the
gravitational lensing.
\begin{figure}[ht]
\begin{center}
\includegraphics[width=6cm]{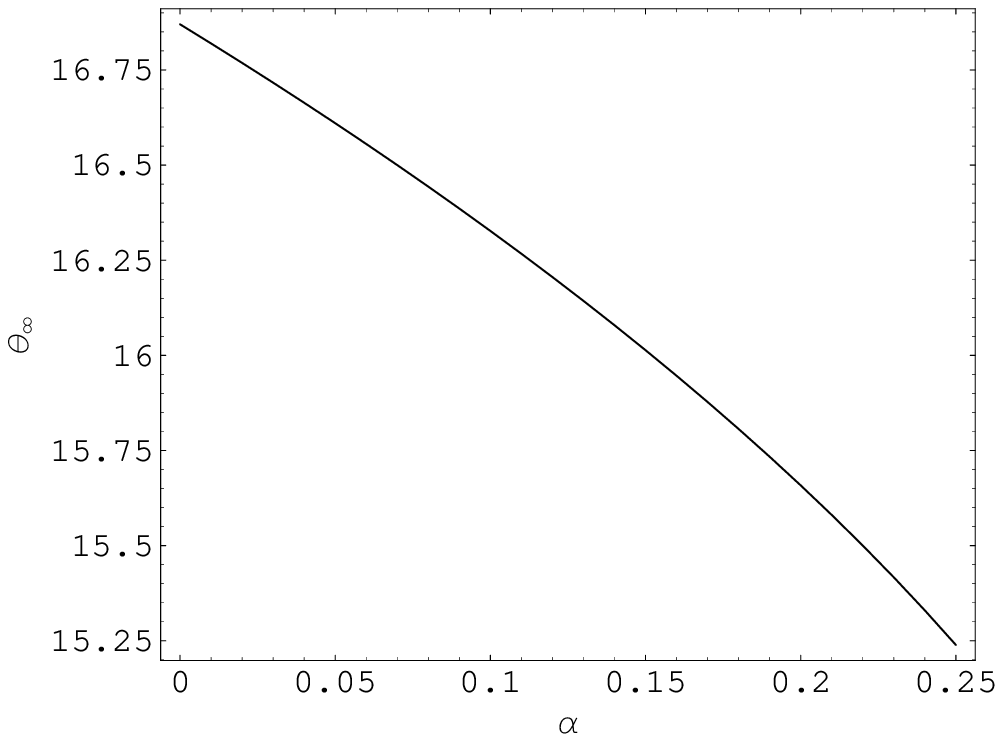}\;\;\;\;\;
 \includegraphics[width=5.9cm]{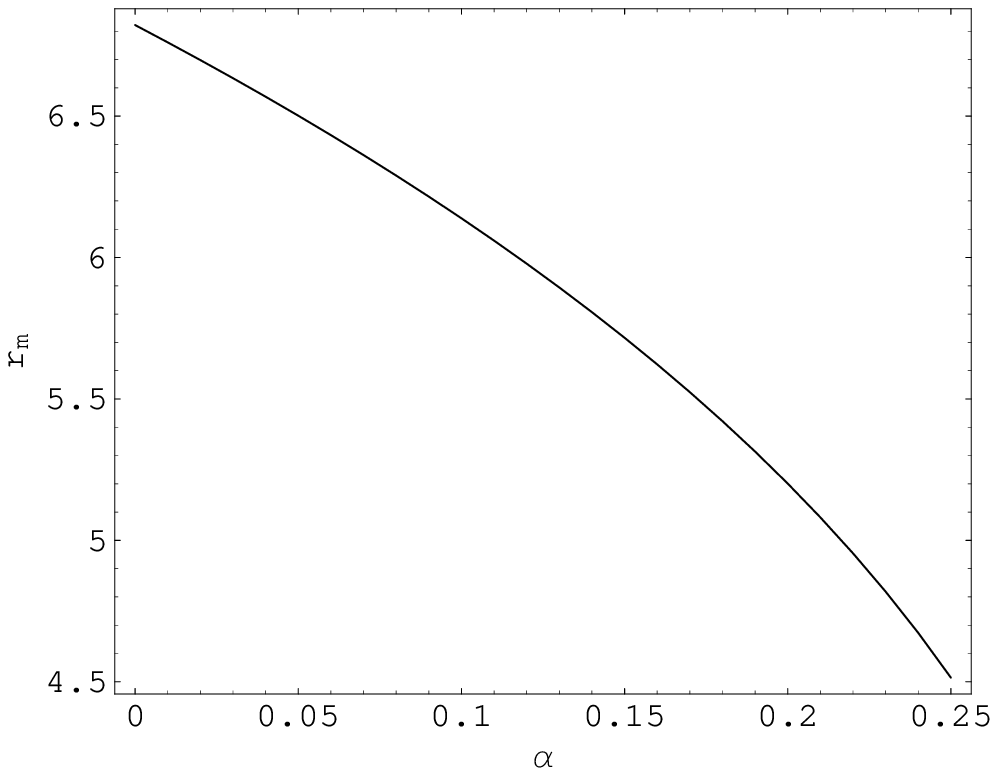}\\
 \includegraphics[width=5.9cm]{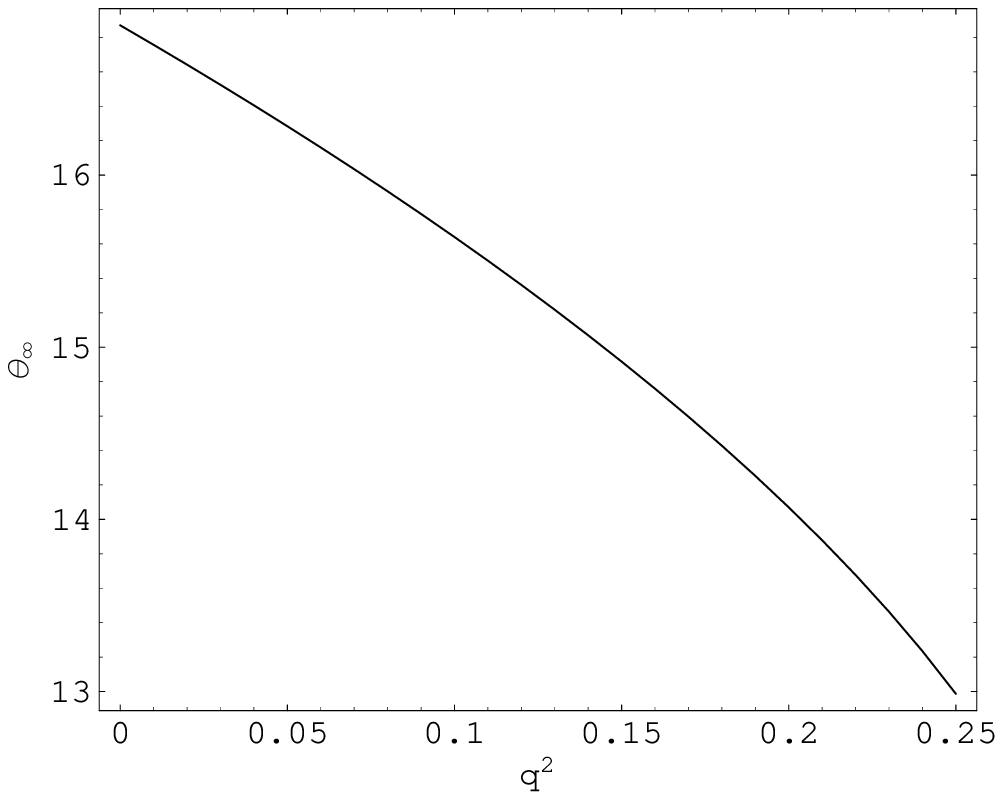}\;\;\;\;\;
 \includegraphics[width=5.9cm]{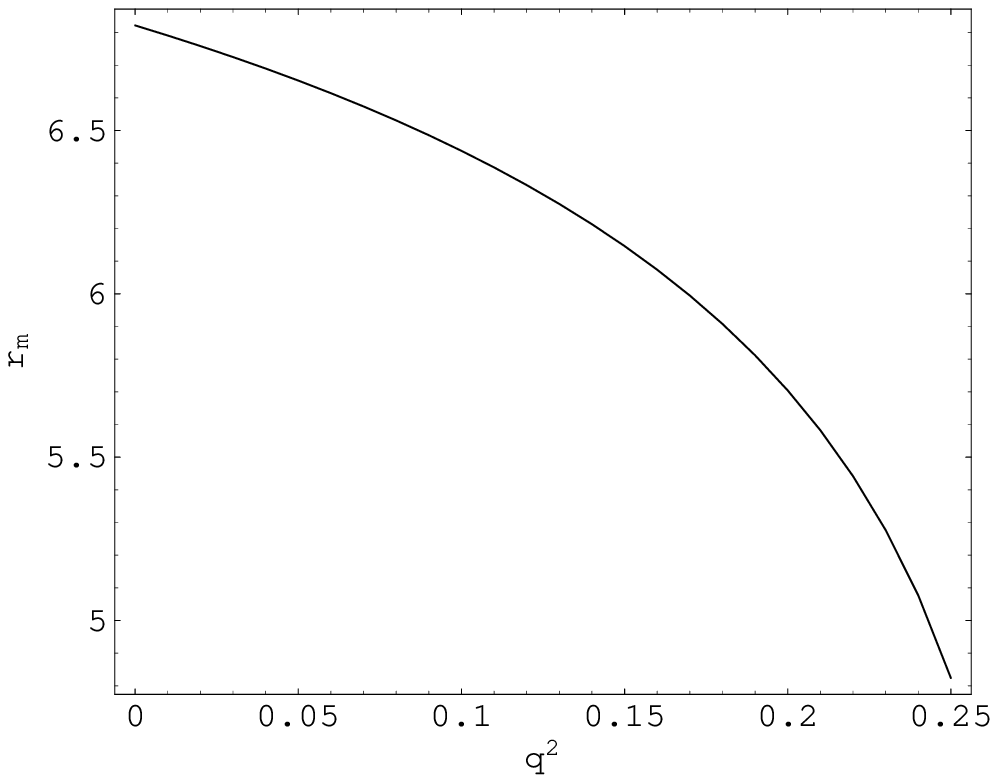}
\caption{Gravitational lensing by the Galactic center black hole.
Variation of the values of the angular position $\theta_{\infty}$
and the relative magnitudes $r_m$ with parameter $\alpha$ in the
deformed Ho\v{r}ava-Lifshitz black hole spacetime (in the upper row)
and with $q^2$ in the Reissner-Norstr\"{om} black hole (in the lower
row).}
 \end{center}
 \label{f2}
 \end{figure}

The mass of the central object of our Galaxy is estimated to be
$2.8\times 10^6M_{\odot}$ and its distance is around $7.6kpc$. For
different $\alpha$, the numerical value for the minimum impact
parameter $u_{ps}$, the angular position of the relativistic images
$\theta_{\infty}$, the angular separation $s$ and the relative
magnification of the outermost relativistic image with the other
relativistic images $r_{m}$ are listed in the table I.
\begin{table}[h]
\begin{center}
\begin{tabular}{|c|c|c|c|c|c|c|}
\hline \hline $\alpha$ &$\theta_{\infty} $($\mu$arcsecs)&\; $s$
($\mu$arcsecs) \;\; & $r_m$(magnitudes)
&\;\;\;\;$u_{ps}/R_S$\;\;\;\; &
\;\;\;\;\;\;\;\;$\bar{a}$\;\;\;\;\;\;\;\; &\;\;\;\;\;\;\;\;
$\bar{b}$\;\;\;\;\;\;\;\; \\
\hline
 0& 16.870& 0.0211& 6.8219&2.598& 1.000& -0.4002 \\
 \hline
0.05&16.610&0.0273&6.5014&2.558& 1.049&-0.4450\\
\hline
0.10& 16.327&0.0368&6.1387&2.514&1.111&-0.4888 \\
\hline
0.15& 16.014&0.0530&5.7162&2.466&1.193&-0.5331 \\
 \hline
0.20 &15.658&0.0835&5.2008&2.411&1.312& -0.5826\\
 \hline
0.25&15.239&0.1541&4.5145&2.347&1.511&-0.6586
 \\
\hline\hline
\end{tabular}
\end{center}
\label{tab1} \caption{Numerical estimation for main observables and
the strong field limit coefficients for black hole at the center of
our galaxy, which is supposed to be described by the deformed
Ho\v{r}ava-Lifshitz black hole spacetime. $\alpha$ is the parameter
of metric. $R_s$ is Schwarzschild radius.
$r_m=2.5\log{\mathcal{R}}$}.
\end{table}
It is easy to obtain that our results reduce to those in the
Schwarzschild black hole sacetime as $\alpha=0$. Moreover, from the
table I, we also find that as the parameter $\alpha$ increases, the
minimum impact parameter $u_{ps}$, the angular position of the
relativistic images $\theta_{\infty}$ and the relative magnitudes
$r_m$ decrease, but the angular separation $s$ increases.

From Fig. (2), we find that in the deformed Ho\v{r}ava-Lifshitz
black hole with the increase of parameter $\alpha$, the angular
position $\theta_{\infty}$ decreases more slowly and $r_m$ more
quickly, but angular separation $s$ increases more rapidly. This
means that the bending angle and the relative magnification of the
outermost relativistic image with the other relativistic images are
smaller in the deformed Ho\v{r}ava-Lifshitz black hole. Our results
also agree with that obtained by in the weak field limit
\cite{RAK1}. In order to identify the nature of these two compact
objects lensing, it is necessary for us to measure angular
separation $s$ and the relative magnification $r_m$ in the
astronomical observations. Tables I tell us that the resolution of
two extremely faint images separated is $\sim 0.06$ $\mu$arc sec,
which is too small. However, with the development of technology, the
effects of parameter $\alpha$ on gravitational lensing may be
detected in the future.

\section{summary}

Gravitational lensing in strong field limit provides a potentially
powerful tool to identify the nature of black holes in the different
gravity theories. In this paper we have investigated strong field
lensing in the deformed Ho\v{r}ava-Lifshitz black hole spacetime.
The model was applied to the supermassive black hole in the Galactic
center. Our results show that with the increase of the parameter
$\alpha$ the minimum impact parameter $u_{ps}$, the angular position
of the relativistic images $\theta_{\infty}$ and the relative
magnitudes $r_m$ decrease. The angular separation $s$ increases.
Comparing with the Reissner-Norstr\"{om} black hole, we find with
the increase of parameter $\alpha$, the angular position
$\theta_{\infty}$ decreases more slowly and $r_m$ more quickly, but
angular separation $s$ increases more rapidly. This may offer a way
to distinguish the deformed Ho\v{r}ava-Lifshitz black hole from a
Reissner-Norstr\"{om} black hole by the astronomical instruments in
the future.

\begin{acknowledgments}
This work was partially supported by the National Natural Science
Foundation of China under Grant No.10875041; the Scientific Research
Fund of Hunan Provincial Education Department Grant No.07B043 and
the construct program of key disciplines in Hunan Province. J. L.
Jing's work was partially supported by the National Natural Science
Foundation of China under Grant No.10675045; the FANEDD under Grant
No. 200317; and the Hunan Provincial Natural Science Foundation of
China under Grant No.08JJ3010.
\end{acknowledgments}

\end{document}